\documentclass[]{aa}
\usepackage{graphicx,txfonts,amssymb,natbib}
\authorrunning{C. Fedeli and M. Bartelmann}
\titlerunning
  {Selection effects}

\begin{document}

\title
  {Selection effects on X-ray and strong-lensing clusters in \\ various
    cosmologies}

\author{C. Fedeli\thanks{E-mail: cosimo@ita.uni-heidelberg.de}
  and M. Bartelmann
  \institute
    {Zentrum f\"ur Astronomie, ITA, Universit\"at Heidelberg,
     Albert-\"Uberle-Str. 2, 69120 Heidelberg, Germany}}

\date{\emph{Astronomy \& Astrophysics, submitted}}

\abstract
 {Galaxy clusters are hotter and more X-ray luminous than in quiescence
while they undergo major mergers, which also transiently increase their
strong-lensing efficiency. We use semi-analytic models for both effects
to study how cluster dynamics in different dark-energy models affects
the X-ray selected cluster population and its strong-lensing optical
depth. We find that mergers increase the number of observable X-ray
clusters by factors of a few and considerably broaden their redshift
distribution. Strong-lensing optical depths are increased by a very
similar amount. Quite independent of cosmology, X-ray bright clusters
above a flux limit of $10^{-13.5}\,\mathrm{erg\,s^{-1}\,cm^{-2}}$ produce
$\sim60\%$ of the strong-lensing optical depth, and only $\sim1\%$ above
a flux limit of $10^{-11.5}\,\mathrm{erg\,s^{-1}\,cm^{-2}}$ if mergers are
taken into account.}


\maketitle

\section{Introduction}

Cluster selection by X-ray emission is generally believed to produce
well-defined samples of massive clusters. This is certainly true for
relaxed objects near virial equilibrium, but clusters undergo
substantial evolution during the cosmic epoch which we can overlook.
Numerical simulations demonstrate that temperatures and X-ray
luminosities of the intracluster gas increase by factors of a few for
periods which are comparable to the sound-crossing time while clusters
undergo major mergers. This may lead to a substantial contamination of
X-ray flux-selected cluster samples by less massive, but dynamically
active clusters. This bias needs to be quantified before cosmological
conclusions based on the cluster population can be considered reliable.

Massive and compact galaxy clusters are also efficient strong lenses.
This gives rise to the expectation that strong lensing should be
particularly frequent in X-ray selected cluster samples, and in fact
many X-ray luminous clusters have been found to be strong gravitational
lenses.

However, strong cluster lensing can also be transiently increased by
factors $\lesssim10$ \citep{TO04.1}
during major cluster mergers, on time-scales
comparable to the dynamical cluster time-scale. This may lift relatively
low-mass clusters above the criticality limit for strong lensing which
would otherwise be undercritical. Based on a relatively small sample of
numerically simulated galaxy clusters, \cite{BA96.2} pointed out that
X-ray selection is not guaranteed to select for the most efficient,
strongly-lensing galaxy clusters. Early work on the interplay between
strong lensing statistics and observational selection effects can also be
found in \cite{WU96.1} and \cite{CO99.4}.

Both effects of major mergers, the enhancement of their X-ray
visibility and their strong-lensing efficiency, potentially open a huge
reservoir of clusters which would remain unobservable in quiescence.
The amplitude of this effect must depend on the frequency of major
mergers, and thus on the cosmological model and its parameters.
Specifically, merger rates at fixed redshift depend on the amount of
dark matter and dark energy and its cosmic evolution.

Here, we address the question how X-ray cluster selection may affect
the strong-lensing efficiency of the selected clusters, and what
fraction of the optical depth for strong lensing we can expect to be
produced by galaxy clusters visible above a certain X-ray flux limit.
In addition to the standard $\Lambda$CDM cosmology, we analyse three
cosmological models with dynamical dark energy which predict different
merger histories. Two of them have a small, but finite dark-energy
density in the early universe and are thus expected to host cluster
populations which form earlier in time.

We combine two semi-analytic methods, one derived by \cite{RA02.1}
describing the enhancement of X-ray temperatures and luminosities during
mergers, and another developed by \cite{FE06.1} for calculating
strong-lensing cluster cross sections. Cluster merger histories are
modelled by merger trees planted in the extended Press-Schechter
formalism.

We summarise the model for the enhanced X-ray temperature and
luminosity in the next section and their translation into observable
fluxes in Sect.~3. We describe the application of this model to
simulated cluster populations in Sect.~4, present our results in
Sect.~5 and conclude in Sect.~6.

\section{Luminosity and temperature boost}

When galaxy clusters undergo violent dynamical events, such as
interactions with substantial substructures or mergers with galaxy
groups and clusters of comparable mass, the intra-cluster medium (ICM) is
compressed and heated by ram pressure and shock waves. This results in
an overall enhancement of the mean gas temperature and the X-ray
emissivity due to bremsstrahlung. The dynamics of gas and dark matter in
clusters during major mergers is typically very complicated and usually
studied based on numerical simulations (see e.g. \citealt{PO06.1,PO07.1} for a
recent review and applications).

It will be sufficient for our purposes to model the short-term increases
in temperature and X-ray luminosity in a simplified manner which
captures their important characteristics in a statistically correct way.
Such a simplified model is given by \cite{RA02.1}. There, the authors
derive fitting formulae for the time-dependent increase in average
temperature relative to its unperturbed value of the combined system of
main cluster and merging body during its interaction.

They employ $N$-body cluster simulations combined with adiabatic
hydrodynamics developed and described in earlier work
\citep{RI01.1,RI00.1}. Shocks in the ICM are extremely well resolved in
these simulations, allowing temperature and luminosity increases to be
studied in detail.

\cite{RA02.1} find that the total time interval $\Delta t$ during which
the average temperature or bolometric luminosity of the ICM of the
system are raised above fixed levels $T$ or $L$ is given by
\begin{equation}\label{eqn:fit}
\xi = \sqrt{[(\Gamma - \Gamma_\mathrm{c})^2 - 1](\varepsilon^2-1)} +
\xi_\mathrm{c},
\end{equation}
where $\Gamma$ is the ratio between the quantity in question, $T$ or
$L$, and its unperturbed value $T_0$ or $L_0$. The parameter
$\Gamma_\mathrm{c}$ is related to the maximum value of the boost,
as will be explained below. The time interval is
measured in units of the sound-crossing time $t_\mathrm{sc}$ of the main
cluster body by $\xi \equiv \log(\Delta t/t_\mathrm{sc})$. Assuming
isothermal gas, the sound-crossing time is
\begin{equation} 
t_\mathrm{sc} = \frac{R_\mathrm{v}}{c_\mathrm{s}} =
R_\mathrm{v}\sqrt{\frac{\mu m_\mathrm{p}}{kT}},
\end{equation}
where the virial radius of the dark matter halo of the main cluster,
$R_\mathrm{v}$, is taken as a characteristic dimension, $\mu$ is the
mean molecular weight of the ICM (assumed to be $\mu = 0.59$ throughout
this work) and $m_\mathrm{p}$ is the proton mass.

The merger is characterised by the mass fraction of the secondary
cluster,
\begin{equation}
f = \frac{M_2}{M_1 + M_2}\;,
\end{equation}
which obviously reaches a maximum value of $1/2$ for equal-mass mergers.
The fit parameters $\Gamma_\mathrm{c}$ and $\varepsilon$ are then expressed
as power laws of $f$,
\begin{equation}
\Gamma_\mathrm{c} = 1+A\,f^B\;,\quad
\varepsilon = C\,f^{-D}\;,
\end{equation}
and
\begin{equation}
\xi_\mathrm{c} = E\,\left[ \ln \left( M_1+M_2 \right) - 
F\ln \left( M_1^{1/3} + M_2^{1/3} \right) \right]\;.
\end{equation}
The amplitudes and exponents appearing in the last three equations were
calibrated by \cite{RA02.1} against their simulations. They are
summarised in Table \ref{tab:fit} for the boosts in both the temperature
and the bolometric $X$-ray luminosity.

\begin{table}[t!]
  \caption{Best-fit amplitudes and exponents for the temperature (top
row) and luminosity (second row) boosts during cluster
mergers.}\label{tab:fit}
  \begin{center}
    \begin{tabular}{|l|r|r|r|r|r|r|}
    \hline
      quantity & A & B & C & D & E & F \\
      \hline
      \hline
      temperature &
      $3.98$ & $0.448$ & $0.96$ & $0.539$ &  $3.71$ & $2.81$ \\
      luminosity &
      $8.28$ & $0.659$ & $0.91$ & $0.316$ & $-0.74$ & $3.29$ \\
    \hline
    \end{tabular}
  \end{center}
\end{table}

\begin{figure*}[t!]
\begin{center}
  \includegraphics[width=0.4\hsize]{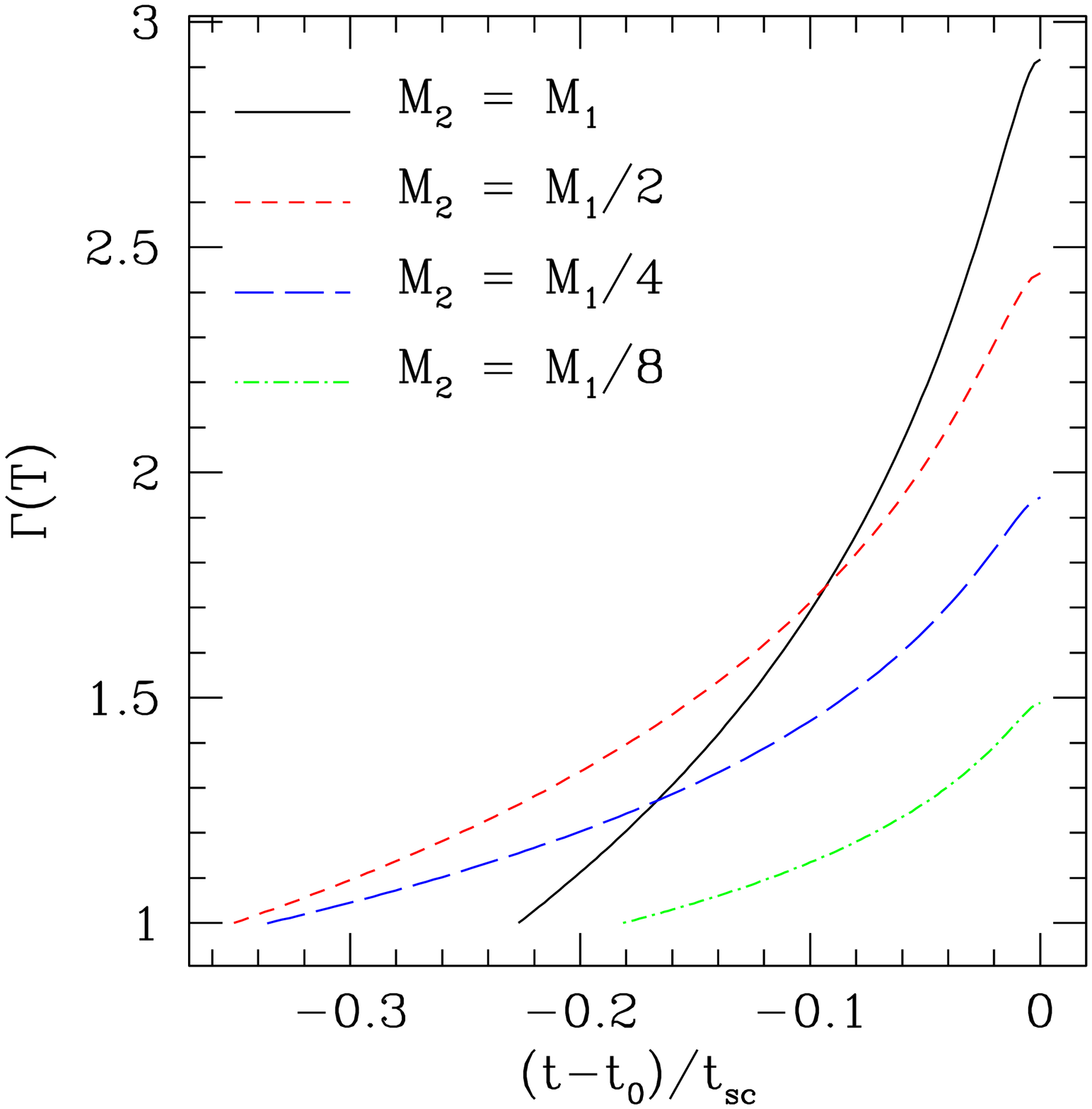}
  \includegraphics[width=0.4\hsize]{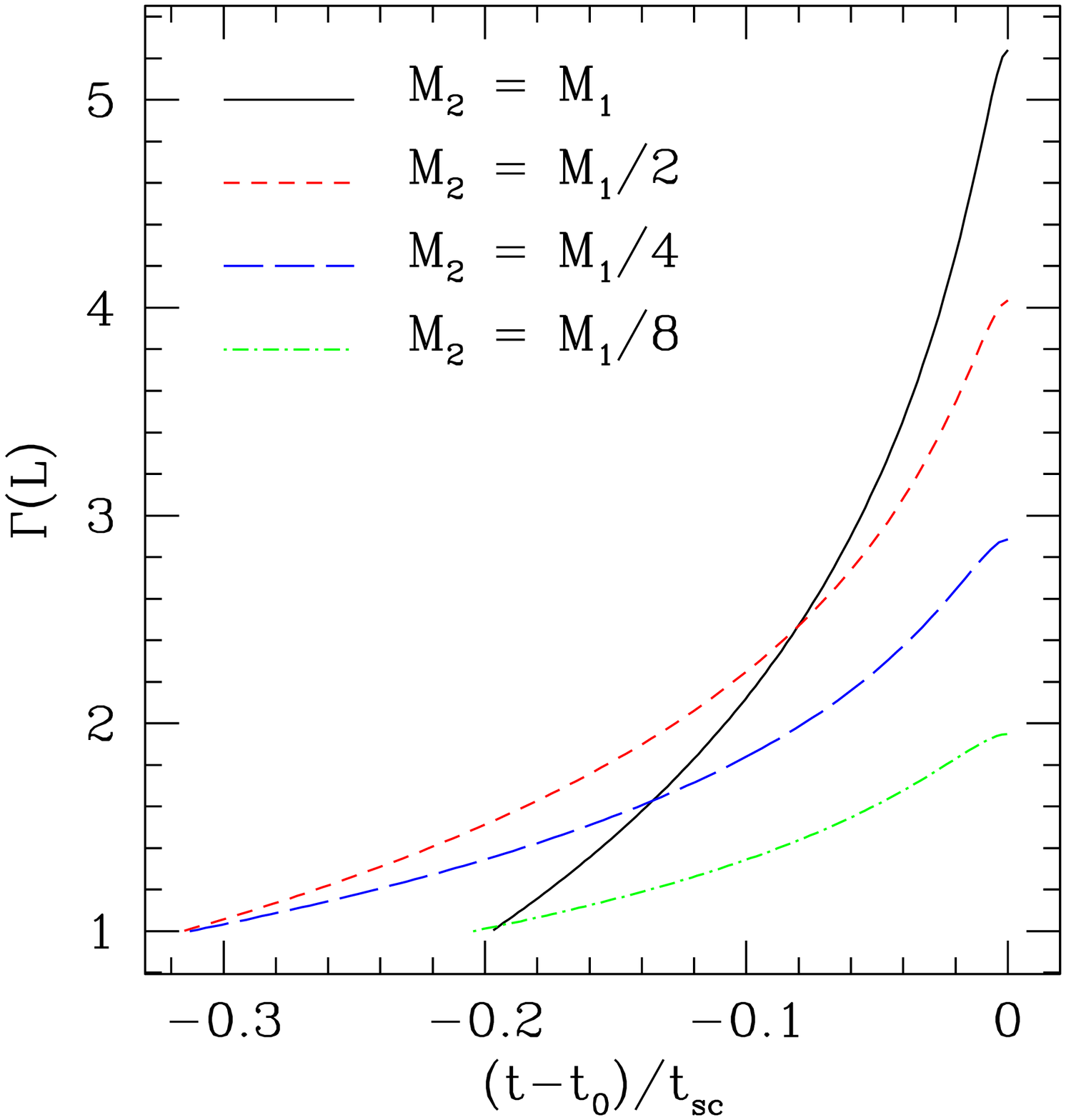}
\end{center}
\caption{The enhancement of temperature (left panel) and luminosity
(right panel) of an interacting system according to \cite{RA02.1}, where
the main cluster has mass $M_1 = 7.5 \cdot 10^{14} M_\odot$, and the
substructure's mass $M_2$ takes four different values, as labelled in the
plot. Time is measured in units of the sound-crossing time of the main
structure and starting from the instant $t_0$ of maximum boost.}
\label{fig:boostFit}
\end{figure*}

Solving Eq.~(\ref{eqn:fit}) for $\Gamma$,
\begin{equation}
\Gamma = \Gamma_\mathrm{c} - \sqrt{1 + \frac{(\xi-\xi_\mathrm{c})^2}
{\varepsilon^2-1}}\;,
\end{equation}
we see that it allows a maximum value for the boost of
$\Gamma_\mathrm{m} = \Gamma_\mathrm{c}-1$.

We plot the function $\Gamma(\xi)$ for the temperature and the
luminosity in Figure \ref{fig:boostFit}. It shows the temperature and
bolometric luminosity of the merging system in units of the pre-merger
values as a function of time in the interval between the beginning of
the boost ($\Gamma = 1$) to the moment of perfect overlap of the two
clusters (and thus of the maximum boost, $\Gamma = \Gamma_\mathrm{m}$).
The mass of the main cluster is set to $M_1 = 7.5 \cdot 10^{14} M_\odot$,
and results are shown for four different values for the mass $M_2$ of
the merging substructure.

As intuitively expected,  these plots show that the maximum temperature
and luminosity reached by the system is larger when the masses involved
in the merger process are similar. In that case, duration of the boost
is also minimal. Moreover, the curves illustrate that the relative
increase in bolometric luminosity exceeds the one in average
temperature, reflecting the higher sensitivity of bremsstrahlung
emission to the density compared to the temperature.

Finally we emphasise that the fitting formulae for the average
temperature and bolometric $X$-ray luminosity enhancements given above
are valid only for head-on mergers. Generalisations to non-zero impact
parameters are given in \cite{RA02.1}, but we assume head-on mergers
throughout for simplicity. Note that, for a non-head on merger,
the duration of the boosts
in temperature and luminosity is larger, but their maximum values are smaller.
These are two somewhat counter-acting effects, and we expect the total
influence to be not significant for our pourposes.

\section{Fluxes obtained from individual clusters}

We shall refer mainly to one particular set of observed galaxy clusters
when comparing to observations, i.e. the ROSAT-ESO Flux Limited X-ray
cluster sample (\textit{Reflex}, \citealt{CO00.1,SC01.1,BO01.1}), which
was drawn from the ROSAT All-Sky Survey (RASS, \citealt{SN90.1}). In
this section, we describe the construction of a synthetic cluster sample
imitating the procedure used for the construction of the \textit{Reflex}
sample.

\subsection{Ideal flux}

We describe the merger history of individual clusters by means of merger
trees constructed based on the extended Press-Schechter theory
\citep{PR74.1,BO91.1,LA93.1,SO99.1}. Thus, the
only information we have on each individual galaxy cluster is its mass
and its redshift. We first related these properties to the
\emph{idealised} X-ray flux, that is the flux that would be measured in
the absence of any instrumental issue. Next, we shall add background
noise, convolution with the point-spread function (PSF), and the
detector response.

We start from the virial relation between mass, redshift and temperature
of the ICM,
\begin{equation}\label{eqn:virial}
kT = 4.88 \mathrm{keV} \left[ \frac{M}{10^{15}M_\odot} h(z)
\right]^{2/3}\;,
\end{equation}
where $h(z)$ is the (reduced) Hubble parameter at the redshift $z$ of
the cluster, and the normalisation constant is calibrated with the
cluster simulations of \cite{MA01.2}.

Introducing the temperature-mass relation (\ref{eqn:virial}) into a
merger tree, we assign temperatures to individual clusters. When a
cluster is merging with a substructure according to its merger tree, we
can either ignore the temperature and luminosity boost caused by the
merger. In this case, only the increasing cluster mass will cause the
temperature to rise. Or, we can boost the temperature according to the
description outlined in the previous section, depending on the state of
the merger process. In both cases, we obtain a unique temperature for
each cluster at each redshift step in its merger tree.

We note here that no statistical fluctuations are taken into account
in our assignment of X-ray temperatures and luminosities to clusters of a
given mass. Thus, once the mass of the main cluster body and a merging
subclump are fixed, the same merger phase will always lead to the same
temperature increase. However, mergers do introduce statistical fluctuations
into the temperature-luminosity relations of our simulated clusters. We shall
return to this point further below.

We also clarify that we consider only binary mergers here. Whenever a
cluster undergoes a multiple merger, we model only the one with the most
massive substructure, neglecting the others in comparison. Since the
simultaneous interaction of a galaxy cluster with more than one massive
substructure is an extremely rare event, we believe that this approximation
is sufficiently accurate for our pourposes.

\begin{figure}[t!]
  \includegraphics[width=1.0\hsize]{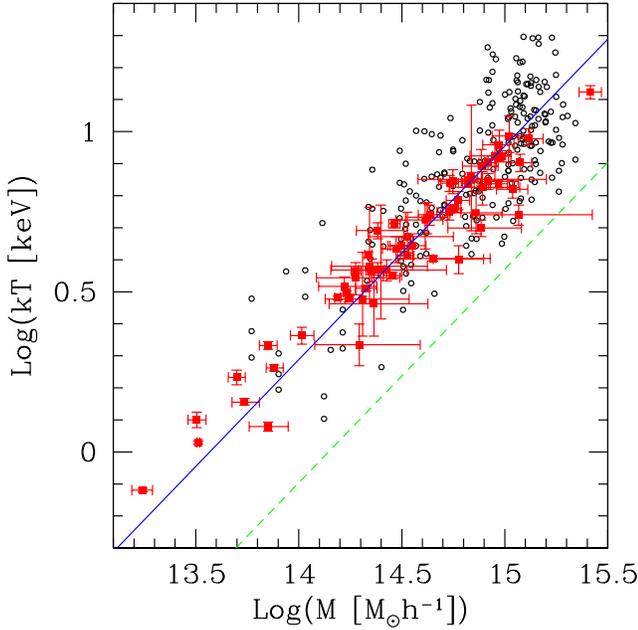}
  \caption{The mass-temperature relation of our synthetic cluster sample.
      Black, empty circles represent our sample clusters at low redshift.
      The green dashed line is the original $M-T$ relation given by
      Eq. (\ref{eqn:virial}). The red filled squares with errorbars are the
      observed clusters of the HIFLUGCS sample, and the blue solid line is a
      relation with the same slope as Eq. (\ref{eqn:virial}) but with a
      normalization higher of a factor $\sim 2$.}
\label{fig:mtl}
\end{figure}

Figure~\ref{fig:mtl} shows the relation between the mass and the
temperature for low-redshift model clusters in our synthetic sample, after
the temperature boost due to mergers has been applied. The
normalization is obviously higher (a factor $\sim 2$) than that of the
original relation Eq.~(\ref{eqn:virial}) because cluster mergers always
increase the cluster temperatures. However, the slope of the relation is well
preserved, and the resulting sample fairly reproduces the observed
HIFLUGCS cluster sample \citep{RE02.1}, which mostly contains low-redshift
clusters. This supports the validity of our model.

Next, we need to derive the ideal flux from the mass, the redshift and
the temperature. We do so using the publically available software
package \texttt{xspec} \citep{AR96.1}, assuming that the ICM can be
described by a Raymond-Smith \citep{RA77.1} plasma model. We set the
metal abundance to $Z = 0.3 Z_\odot$ (\citealt{FU98.1,SC99.2}, see also
\citealt{BA03.4}). We normalise the spectrum by means of the
observationally calibrated relation
\begin{equation}\label{eqn:obs}
L_\mathrm{bol} = 2.5 \cdot 10^{43} \mathrm{erg}\,\mathrm{s}^{-1} h^{-2}
\left( \frac{kT}{1.66 \mathrm{keV}} \right)^{2.331}, 
\end{equation}
derived by \cite{AL98.2}, where $L_\mathrm{bol}$ is the bolometric
$X$-ray luminosity (see also  \cite{MU97.1,RE99.2,HA02.2} for some
discussion on the redshift evolution of this $L_\mathrm{bol}-T$
relation). 

Several authors \citep{ST06.1,OH06.1,PR06.1} have discussed that
the observed scatter in the
temperature-mass and luminosity-temperature relations might
not be entirely caused by recent mergers, but rather be sensitive
to the complete merger history of the cluster. 
However, we verified that the scatter in the natural logarithm of mass
around the best fit mass-luminosity relation that we obtain for nearby
clusters in our synthetic sample is $\sigma_{\mathrm{ln} M} \simeq 0.4$
and thus agrees well with the observed value for the HIFLUGCS data
\citep{RE02.1,ST06.1}. Hence, we conclude that our modeling of mergers
introduces scatter into the luminosity-mass relation compatible with the
observed scatter, and thus fairly captures the observed statistical
fluctuations.

Finally, we take account of the interstellar absorption by neutral
hydrogen in the Milky Way. We do so by combining our Raymond-Smith
plasma model with the \texttt{phabs} multiplicative model component of
the \texttt{xspec} software, adopting a constant hydrogen column density
of $n_H = 4 \cdot 10^{20}$ cm$^{-2}$, appropriate for relatively high
galactic latitudes \citep{DI90.1}.

\subsection{Instrumental effects}

In constructing the \textit{Reflex}
cluster sample, \cite{BO01.1} used the count
rate received for each individual cluster by the ROSAT PSPC detector in
the energy channels covering the [0.5, 2.0] keV energy band. To compute
synthetic count rates for each object in our simulated cluster
population, we first need to introduce a model for the distribution of
the ICM within the clusters. We adopt the isothermal $\beta$-model
\citep{CA76.1,CA78.1}, assuming $\beta = 2/3$ throughout, following
\cite{MO99.1}.

The resulting gas-density profile is
\begin{equation}
\rho_\mathrm{ICM}(r) = \frac{\rho_{\mathrm{ICM},0}}{1 +
r^2/r_\mathrm{c}^2}.
\end{equation}
Its core radius $r_\mathrm{c}$ is related to the X-ray
luminosity in the [0.5,2.4] keV energy band through
\begin{equation}
r_\mathrm{c} = 0.125 \mathrm{Mpc}\,h^{-1} \left( \frac{L}{5 \cdot 10^{44}
\mathrm{erg s}^{-1}} \right)^{0.2}
\end{equation}
(\citealt{JO98.1}; see \citealt{VI02.1} for a discussion on the redshift
evolution of this relation).

Since the emissivity of thermal bremsstrahlung is proportional to the
squared gas density, this density profile implies the surface brightness
profile
\begin{equation}
B(\theta) = \frac{B_0}{\left(1 + \theta^2/\theta_\mathrm{c}^2
\right)^{3/2}},
\end{equation}
normalised by
\begin{equation}
B_0 = \frac{F}{2 \pi \theta_\mathrm{c}^2}.
\end{equation}
Here, $F$ is the total flux in the respective energy band (i.e.~the
integral of the surface brightness profile over the solid angle),
$\theta = r/D_\mathrm{A}(z)$ and $\theta_\mathrm{c} =
r_\mathrm{c}/D_\mathrm{A}(z)$. Obviously, $D_\mathrm{A}(z)$ is the
angular-diameter distance to redshift $z$.

We next convolve this surface-brightness profile with the instrumental
PSF. The shape of the ROSAT-PSPC PSF is summarised in \cite{BA03.4}
based on \cite{HA95.1}. Its shape depends slightly both on the energy
channel considered and on the off-axis angle of the source. For
simplicity, we shall assume on-axis sources and an energy channel at 1
keV, approximately at the centre of the energy bands considered in the
present work.

Background count rates are provided in form of a map on the RASS web
page\footnote{\texttt{http://www.xray.mpe.mpg.de/cgi-bin/rosat/}
\texttt{rosat-survey}}.
We use a constant median value of $b = 2.6 \cdot
10^{-4}$ s$^{-1}$ arcmin$^{-2}$, but note that our results are quite
independent of the background correction.

Finally, we obtain the count rate produced by the PSF-convolved,
background-corrected surface-brightness profile. We integrate over the
complete profile using the \texttt{fakeit} command of the \texttt{xspec}
software, adopting the PSPC response matrix in the [0.5,2.0] keV energy band.

\begin{figure}[t!]
  \includegraphics[width=1.0\hsize]{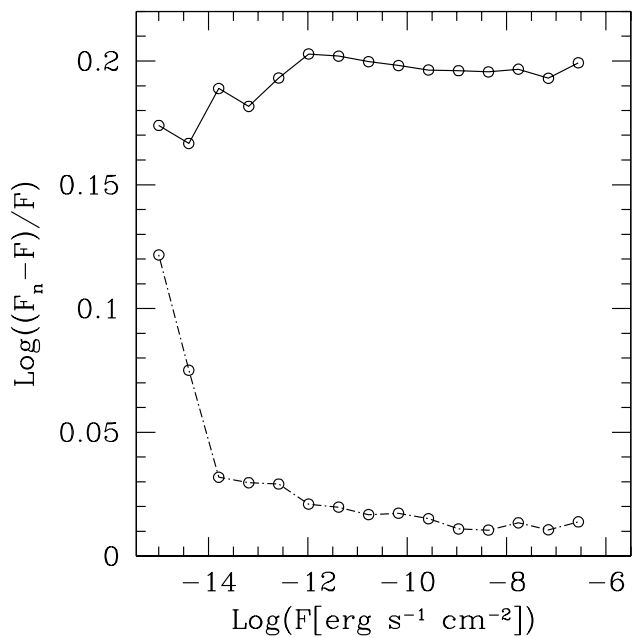}
  \caption{The normalized difference between the nominal and the ideal
    flux as a function of the ideal flux itself is shown for the synthetic
    cluster population used in this work. The upper solid and lower
    dash-dotted curves show the mean and the scatter, respectively. 
    The nominal flux is always $\sim 20\%$ larger than the ideal flux, while
    the scatter in the relation can be neglected only for high fluxes.}
\label{fig:nomId}
\end{figure}

\subsection{Nominal flux}

The nominal flux $F_\mathrm{n}$ for the \textit{Reflex} cluster sample
is defined as the flux produced in the [0.1,2.4] keV energy band by a
Raymond-Smith model plasma set to redshift zero, with a temperature of 5
keV, metal abundance of $Z = 0.3 Z_\odot$, absorption as given by
\cite{DI90.1} and a spectrum normalised so as to reproduce the observed
number counts in the energy channels corresponding to the [0.5,2.0] keV
band \citep{CO00.1,BO01.1}.

To each cluster and at each redshift step of its merger tree, we assign
a nominal flux exactly in the same way. The only difference between the
definitions of our synthetic sample and of the \textit{Reflex} sample is
that we normalise the spectrum of the plasma model so as to reproduce
the count rates computed at the end of Subsect.~3.2.

The \textit{Reflex} cluster sample is flux-limited, in the sense that it
contains only galaxy clusters with nominal flux $\le F_\mathrm{n,lim} =
3.0 \cdot 10^{-12}$ erg s$^{-1}$ cm$^{-2}$. We here adopt the same nominal
flux as a threshold for synthetic cluster samples. We shall use the
nominal \textit{Reflex} flux limit and four additional lower flux limits
in order to create synthetic samples containing a larger number of
objects.

The relation between this nominal flux and the
ideal flux introduced in Sect.~3.1 (without hydrogen absorption)
is shown in Fig.~\ref{fig:nomId}, where the
mean difference between the two fluxes, normalised to the ideal flux, 
and its standard deviation are plotted
as functions of the ideal flux itself.

According to the definition given at the beginning of this section,
the ideal flux of a cluster is computed in the energy band [0.5,2.0] keV,
while the nominal flux is the flux in the [0.1,2.4] keV band of a fiducial
cluster with fixed physical properties ($Z=0.3 Z_\odot$, $T=5$keV, $z=0$) that
produces the same count rates as the cluster at hand in the [0.5,2.0] keV band.
Since the nominal flux is computed in a wider and softer band than the ideal
flux, there is a bias
because the nominal flux exceeds the ideal flux typically by $\sim 20\%$.
The scatter about the mean is relatively large for small fluxes,
but drops to zero as the flux increases because then the effects of
PSF convolution and background  subtraction are smaller. We conclude from this
plot that the ideal may be used instead of the nominal flux, thus saving the
time for the computation of hydrogen absorption and instrumental effect, but
only when the flux is sufficiently large ($\gtrsim 10^{-12}$ erg
s$^{-1}$ cm$^{-2}$) and accounting for the $20\%$ bias.

\section{Application to the cluster population}

\subsection{Cosmological models}

We base our analysis on the same cosmological models analysed in
\cite{FE07.1}. These are a standard $\Lambda$CDM model, a universe with
a constant equation-of-state parameter $w_\mathrm{de} = -0.8$ for the
dark energy, and two early-dark energy models
\citep{FE98.1,DO01.1,DO01.2,CA03.2,WE04.1}. In such models, the
quintessence energy density tracks that of the dominant component of the
cosmic fluid until it begins dominating in the late cosmic history. They
are characterised by a small but non-negligible amount of dark energy at
very early times, which may influence all kinds of high-redshift
processes. \cite{BA06.1} showed that early dark energy also affects the
non-linear structure formation, giving rise to significant differences
in the mass function and the merger rates of dark-matter halos compared
to the concordance $\Lambda$CDM model.

The matter power spectrum is scale-invariant ($n=1$) for the models with
a constant $w_\mathrm{de}$, while the spectral indices are $n = 1.05$ in
the first and $n = 0.99$ in the second early-dark energy model
(hereafter called models EDE1 and EDE2 respectively). The other
cosmological parameters for all models studied here are summarised in
Tab.~\ref{tab:cos}. They were set such as to match the power spectrum of
the CMB temperature fluctuations \citep{SP03.1,SP06.1}, constraints from
the large-scale structure of the Universe \citep{TE04.1}, and
observations of type-Ia supernovae \citep{RI04.1}. Note that this
requires different normalisations $\sigma_8$ for the dark-matter power
spectrum in the different cosmological models, which also affect the
differences in the mass-assembly histories. \cite{FE07.1} studied the
total strong-lensing optical depth of the cluster population expected in
these model cosmologies. Here, we extend this study by taking X-ray
selection effects into account and exploring the effect of early dark
energy on the number of clusters in flux-limited samples.

\begin{table}[t!]
  \caption{Parameters of the four cosmological models studied here.}
  \label{tab:cos}
  \begin{center}
    \begin{tabular}{|l|l|l|l|l|}
      \hline
      & EDE1 & EDE2 & $w_\mathrm{de} = -0.8$ & $\Lambda$CDM\\
      \hline
      \hline
      $\sigma_8$ & 0.82 & 0.78 & 0.80 & 0.84\\
      $h$ & 0.67 & 0.62 & 0.65 & 0.65\\
      $\Omega_{\mathrm{m},0}$ & 0.33 & 0.36 & 0.30 & 0.30\\
      $\Omega_{\mathrm{de},0}$ & 0.67 & 0.64 & 0.70 & 0.70\\
      \hline
    \end{tabular}
  \end{center}
\end{table}

\subsection{Merger trees}

As already stated, we use here the merger trees produced in
\cite{FE07.1} for a set of $N =
500$ dark matter haloes with masses uniformly distributed between
$10^{14} M_\odot h^{-1}$ and $2.5 \cdot 10^{15} M_\odot h^{-1}$. In
\cite{FE07.1}, we computed for each cluster the lensing cross section
for gravitational arcs with length-to-width ratio larger than $d=7.5$,
using the fast, semi-analytic method developed by \citealt{FE06.1}. The
objects in the sample were evolved up to a source redshift randomly
drawn from an observed distribution of faint blue background galaxies
\citep{SM95.1,BA01.1}. Lensing efficiencies were calculated both ignoring and
accounting for mergers with substructures which transiently enhance the
cross sections.

We apply the same scheme here, calculating the nominal X-ray flux for
each sample cluster at each redshift step. We do that both ignoring and
accounting for the effect of mergers which transiently enhance the
intrinsic luminosity and temperature (and thus also the nominal flux) of
the clusters.

Finally, we can combine this information with the calculations of the
lensing efficiencies, evaluating the effect of flux selection on the
statistics of gravitational arcs in galaxy clusters.

\subsection{Cluster number counts}

\begin{figure}[t!]
  \includegraphics[width=1.0\hsize]{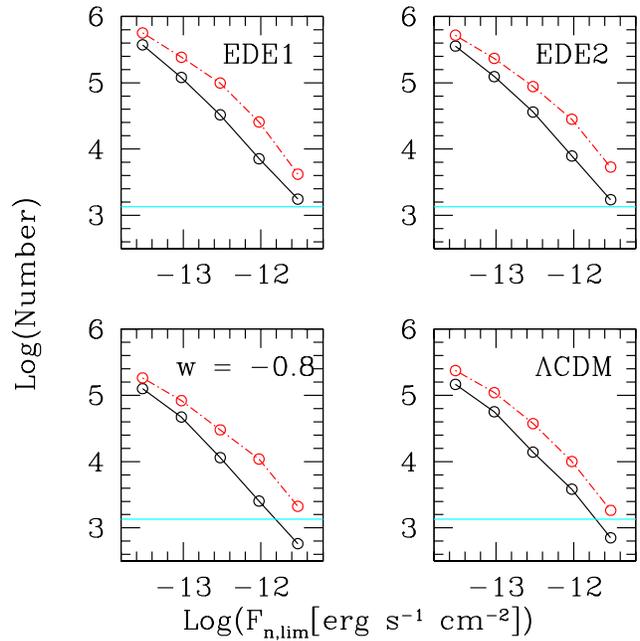}
\caption{The total number of galaxy clusters observable on the whole
sky, given as a function of the limiting nominal flux for the four
different cosmological models considered here. Black solid lines are
obtained by ignoring the transient boost due to cluster mergers, red
lines are obtained taking it into account. The cyan horizontal line
gives the number of galaxy clusters obtained from the \textit{Reflex}
cluster sample extrapolated to the whole sky ($F_\mathrm{n,lim} = 3 \cdot
10^{-12}$ erg s$^{-1}$ cm$^{-2}$).}
\label{fig:nCounts}
\end{figure}

\begin{figure}[ht!]
  \includegraphics[width=1.0\hsize]{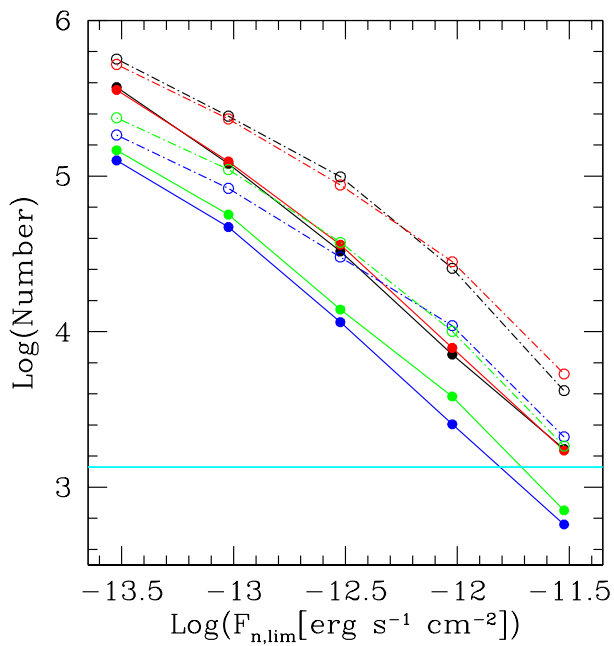}
\caption{As Fig.~\ref{fig:nCounts}, but combining all curves in the same
plot. Black and red curves are for the EDE1 and EDE2 models,
respectively. The blue curve is for the model with constant
equation-of-state parameter $w_\mathrm{de} = -0.8$, and the green line
is for the $\Lambda$CDM model. For each model, the solid and dashed
curves are obtained ignoring mergers and taking them into account,
respectively. The horizontal line shows the number of clusters observed
in the \textit{Reflex} cluster sample.}
\label{fig:nComp}
\end{figure}

\begin{figure}[ht]
  \includegraphics[width=1.0\hsize]{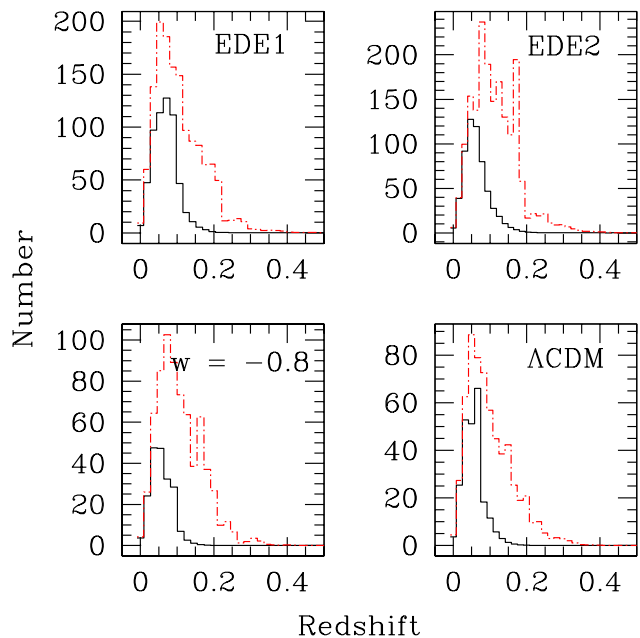}
\caption{Histograms for the cluster redshift distributions expected in
the cosmological models studied here, as labelled in the boxes. The
limiting flux is the same as for the \textit{Reflex} cluster sample,
$F_\mathrm{n,lim} = 3 \cdot 10^{-12}$ erg s$^{-1}$ cm$^{-2}$. As before,
the black lower curve does not account for cluster mergers, while the
red upper curves do.}
\label{fig:zHist}
\end{figure}

It is interesting note in passing how the previous calculations predict
the total number of clusters in a flux-limited sample to change as a
function of the limiting flux in the different cosmological models, and
what the quantitative effect of mergers is in this respect.

We emphasise here that our algorithm for producing synthetic cluster
samples is not ideally adapted to cluster-abundance studies because our
cluster sample includes only relatively high-mass haloes, and thus the
low-mass end of the distribution is not well sampled. Moreover, we did
not take into account the likely steepening of the
temperature-luminosity relation for low-mass clusters or galaxy groups.
While this has no effect on samples of X-ray luminous, hot or strongly
lensing clusters, it is likely that the overall number of structures in
the different flux-limited samples is overestimated. This is because we
tend to assign to objects with very low mass a temperature higher than
expected in presence of steepening.

In Figs.~\ref{fig:nCounts} and \ref{fig:nComp}, we show the total number
of galaxy clusters predicted to be observed in the four cosmological
models used here as a function of the limiting nominal flux. We
plot results obtained by accounting for and ignoring the effects of
cluster mergers, and indicate the total number of galaxy clusters
observed in the \textit{Reflex} cluster sample, extrapolated to the
whole sky.

Several interesting pieces of information can be read off these figures.
First of all, cluster mergers increase the total number of visible
objects by factors between 2 and 3. This factor tends to decrease
towards lower flux limits because of two effects. First, at low flux
limits, the total number of clusters observable without mergers is
larger, thus the fractional increase due to cluster interactions tends
to be smaller. Second, at low flux limits, we include low-mass objects
into the sample whose merger frequency is lower. We also see that,
according to this analysis, only the models with constant
equation-of-state parameter are in agreement with the \textit{Reflex}
observations, while early-dark energy models overpredict the cluster
abundance by a factor of $\approx 2$.

In Fig.~\ref{fig:zHist}, we fix the nominal flux limit to that of the
\textit{Reflex} sample, $F_\mathrm{n,lim} = 3 \cdot 10^{-12}$ erg s$^{-1}$
cm$^{-2}$ and show a histogram of the redshift distribution of observed
clusters. Again, we show results with and without the enhancements by
cluster mergers. For all models, the number of clusters drops to zero
above $z \approx 0.3$ with mergers, and already above $z \approx 0.15$
without mergers. The absence of substantial differences between
different cosmologies is due to the fact that at low redshift the
difference between the structure formation in presence or absence of
early-dark energy tends to disappear (see also the discussion in
\citealt{FE07.1}). It is interesting to note that the observational
results from the \textit{Reflex} sample \citep{CO00.1} are qualitatively
very well reproduced only accounting for cluster mergers. Ignoring the
effect of interactions in our models leads to an underestimate of
objects in the high-redshift tail.

In the context of Fig.~\ref{fig:zHist}, we also note that while the
qualitative trend and the peak position of the observed \textit{Reflex}
distribution are reproduced, the normalisation is generally too high,
for the reasons discussed.

\section{Results}

We finally return to the main purpose of this work, that is probe how
cluster selection by their X-ray flux may influence the optical depth of
the sample for the production of pronounced gravitational arcs when the
effect of cluster mergers are taken into account. In other words, we
analyse how the total number of arcs that we can expect to observe in an
X-ray selected galaxy-cluster sample depends on the X-ray flux limit of
the sample itself. While we have considered only arcs with $d \ge 7.5$
in \cite{FE07.1}, we extend the analysis here to arcs with $d \ge 10$
(so-called giant arcs).

\begin{figure*}[t!]
  \includegraphics[width=0.45\hsize]{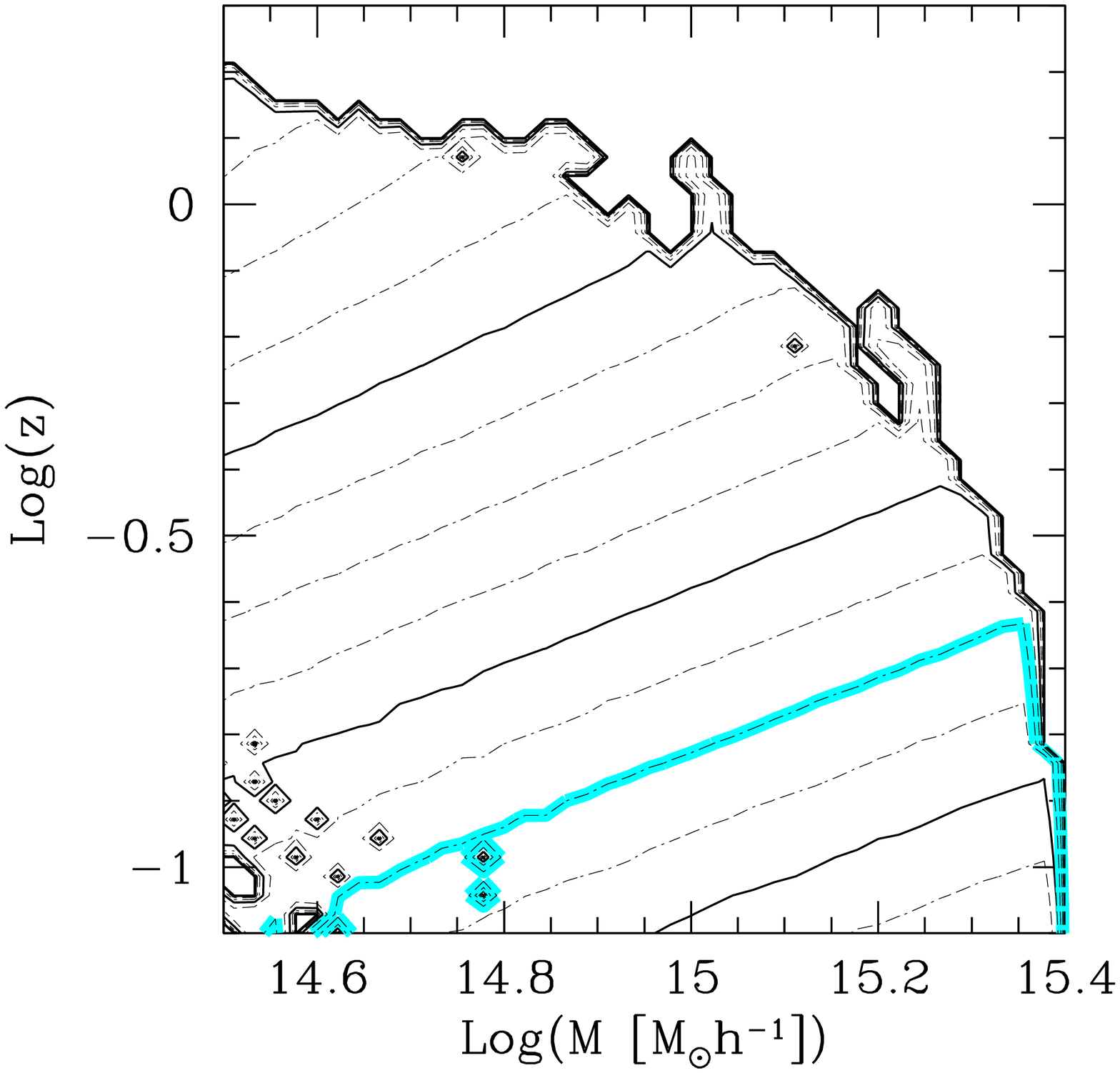}\hfill
  \includegraphics[width=0.45\hsize]{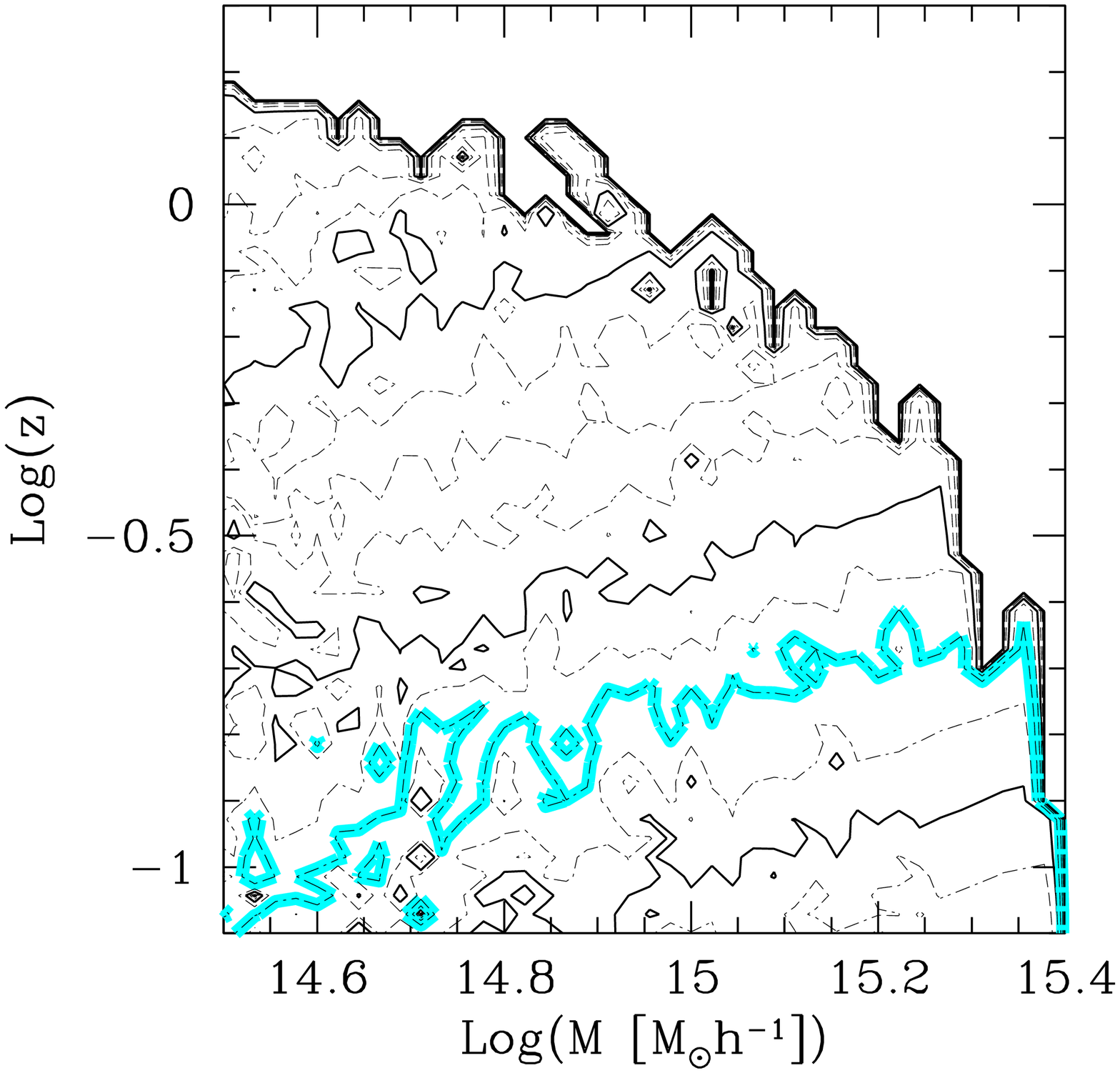}\hfill
  \includegraphics[width=0.45\hsize]{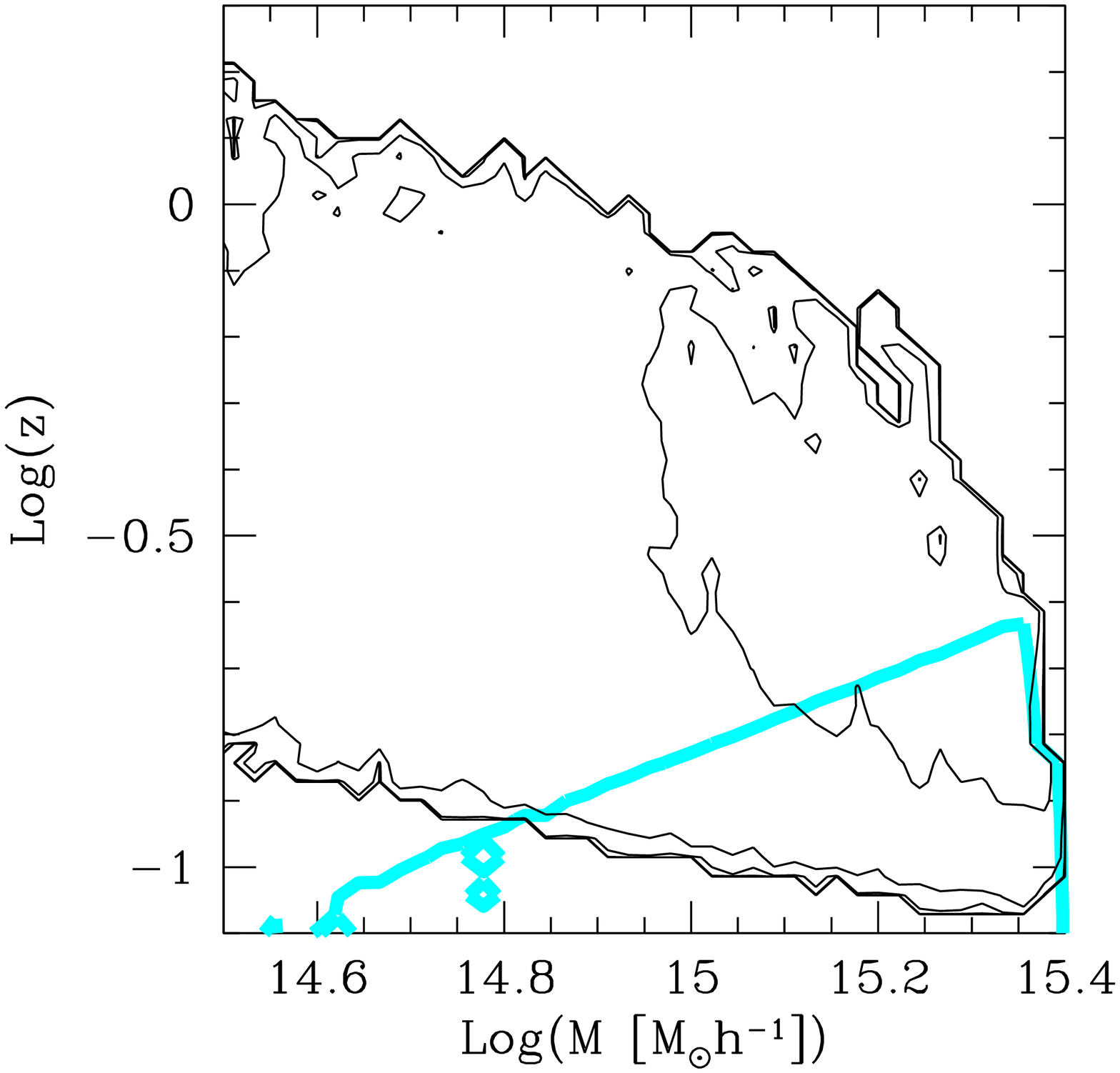}\hfill
  \includegraphics[width=0.45\hsize]{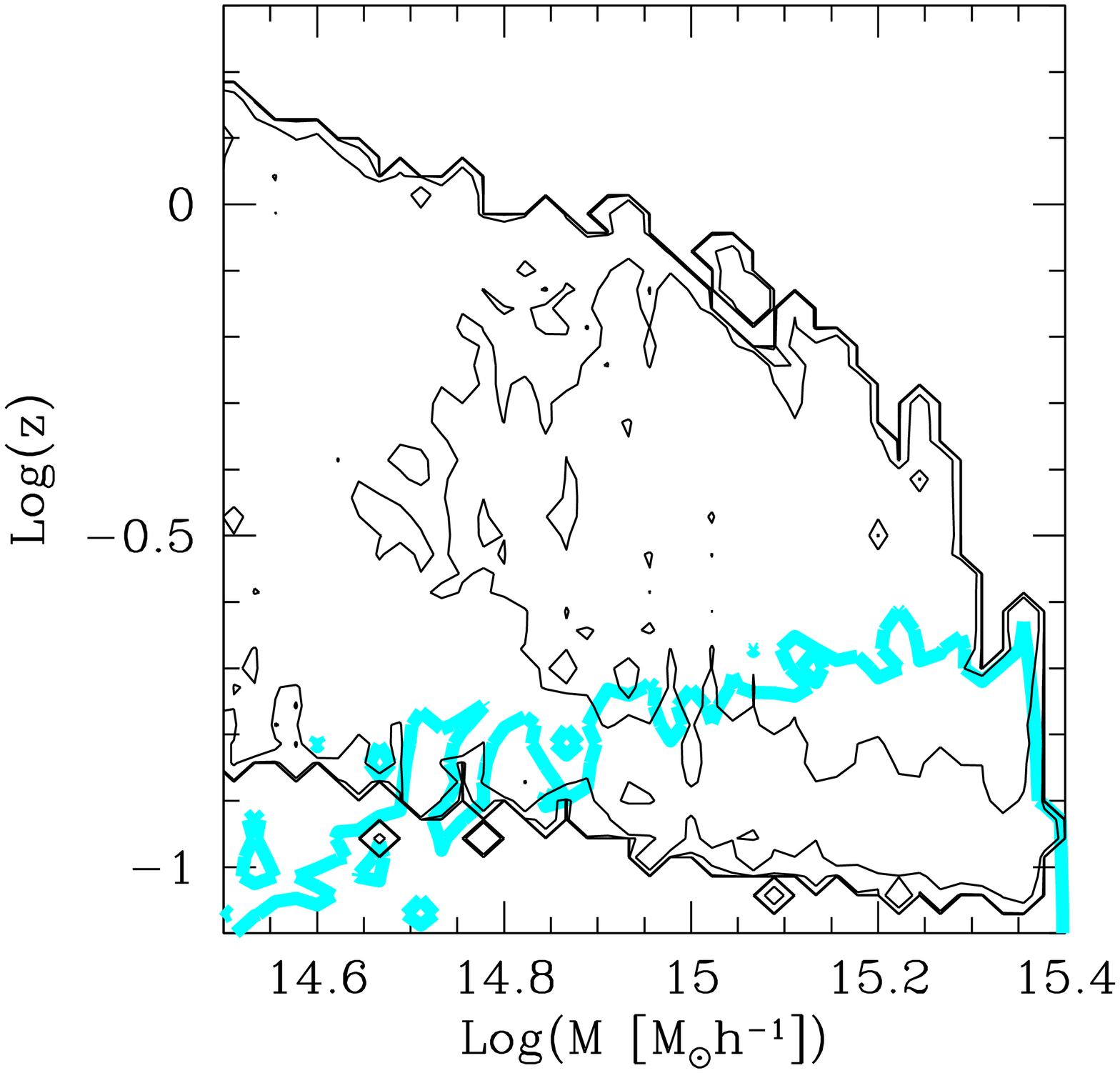}
\caption{Contour lines for the nominal flux (upper panels) and for the
lensing cross section for gravitational arcs with $d \ge 7.5$ (lower panels)
in the $\Lambda$CDM model. The thick cyan contours correspond to the
limiting nominal flux of the \textit{Reflex} cluster sample. The upper
heavy black contour line in the upper panels corresponds to a flux
of $10^{-14}$ erg s$^{-1}$ cm$^{-2}$, the others are spaced by one order
of magnitude.
Analogously, the lowest black contour in the lower panels corresponds to a
cross section of
$10^{-4}$ Mpc$^2 h^{-2}$, and the others are spaced of one order of
magnitude. Merger processes are taken into account in the right panels,
while they are ignored in the left ones.}
\label{fig:mz}
\end{figure*}

Figure~\ref{fig:mz} shows contour lines in the mass-redshift plane for
the nominal X-ray flux of clusters from our synthetic sample, and for
the cross section for arcs with length-to-width ratio larger than $d =
7.5$ for the $\Lambda$CDM cosmological model. Contours in the left and
right panels were obtained ignoring mergers and taking them into
account, respectively. We overplot the flux limit for the
\textit{Reflex} cluster sample. The edge in both the X-ray flux and
lensing-efficiency contour lines going from the upper left to the lower
right corner illustrates the lack of high-mass clusters at high
redshift. These figures clearly show the effect of mergers on the
lensing efficiency and the average X-ray flux from clusters. Including
mergers, the contour lines are much more irregular and extend towards
lower masses and higher redshifts, both for the nominal X-ray flux and
for the cross section. The geometric suppression of the lensing
efficiency at very low redshift is also evident since the black contours in
the lower panels
never reach $z=0$, and turn further away from $z=0$ for lower cluster
masses. The lensing efficiency expected to be observed in a $\Lambda$CDM
model in a \textit{Reflex}-like cluster sample is thus contributed only
by those clusters falling between the cyan curve and the lower
contour lines in the lower panels.

We emphasise that Fig.~\ref{fig:mz} only shows the
properties of individual objects in our synthetic sample. In order to find
sample properties, this
information must be convolved with the cluster mass function. This means
that even the small wiggles appearing in the heavy contour in
Fig.~\ref{fig:mz} when mergers are included will have a substantial effect on
lensing statistics and number counts. 
We prefer not to weight with the mass function in Fig.~\ref{fig:mz}
to illustrate exclusively the effect of the flux cut.
To have an idea of the consequence that small shifts in the $M-z$ plane can
have on the quantitative results, the number of structures in the narrow
mass interval $14.5\le\log M\le14.6$ doubles when computed
for $z \le 0.10$ compared to $z \le 0.13$.

\begin{figure*}[ht]
  \includegraphics[width=0.45\hsize]{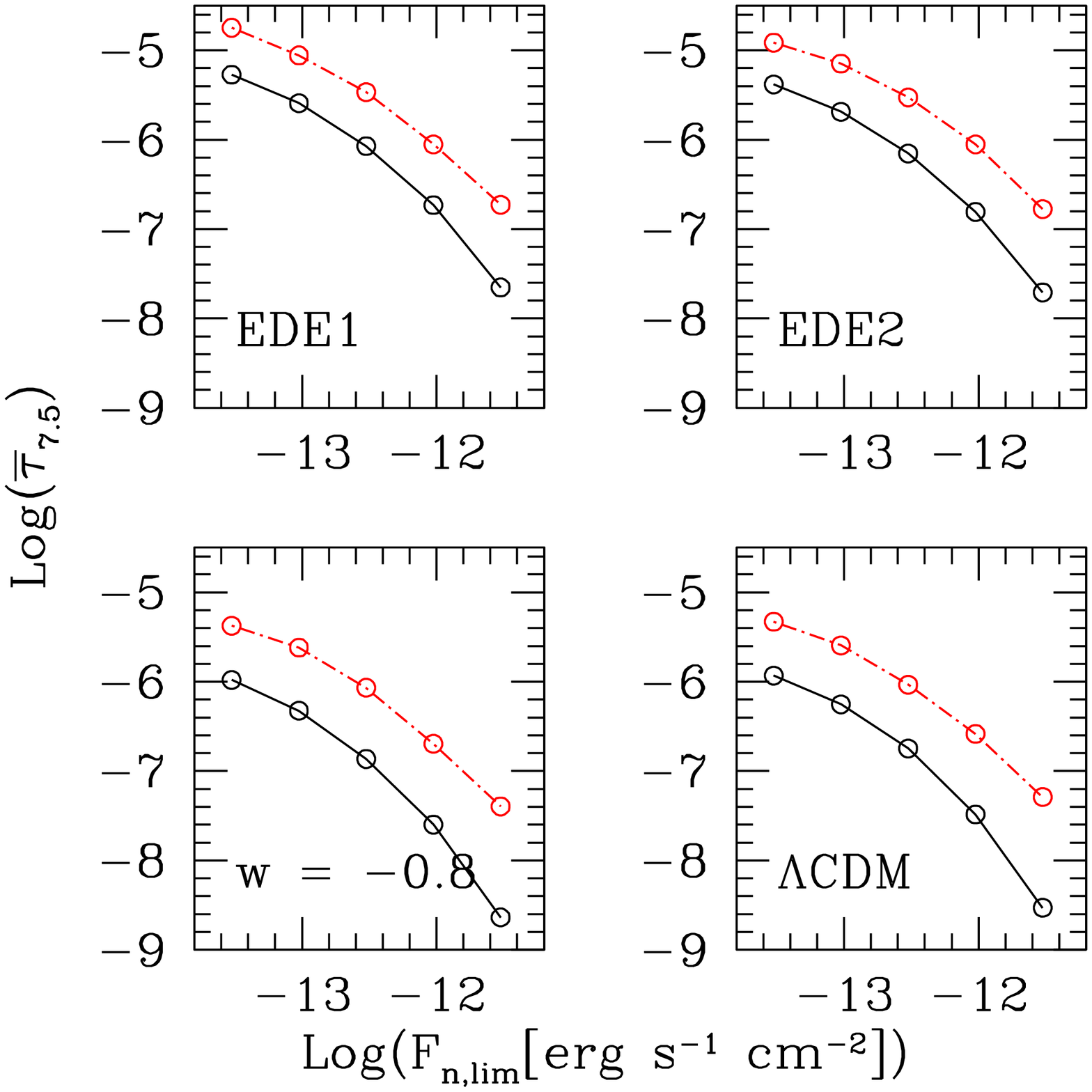}\hfill
  \includegraphics[width=0.45\hsize]{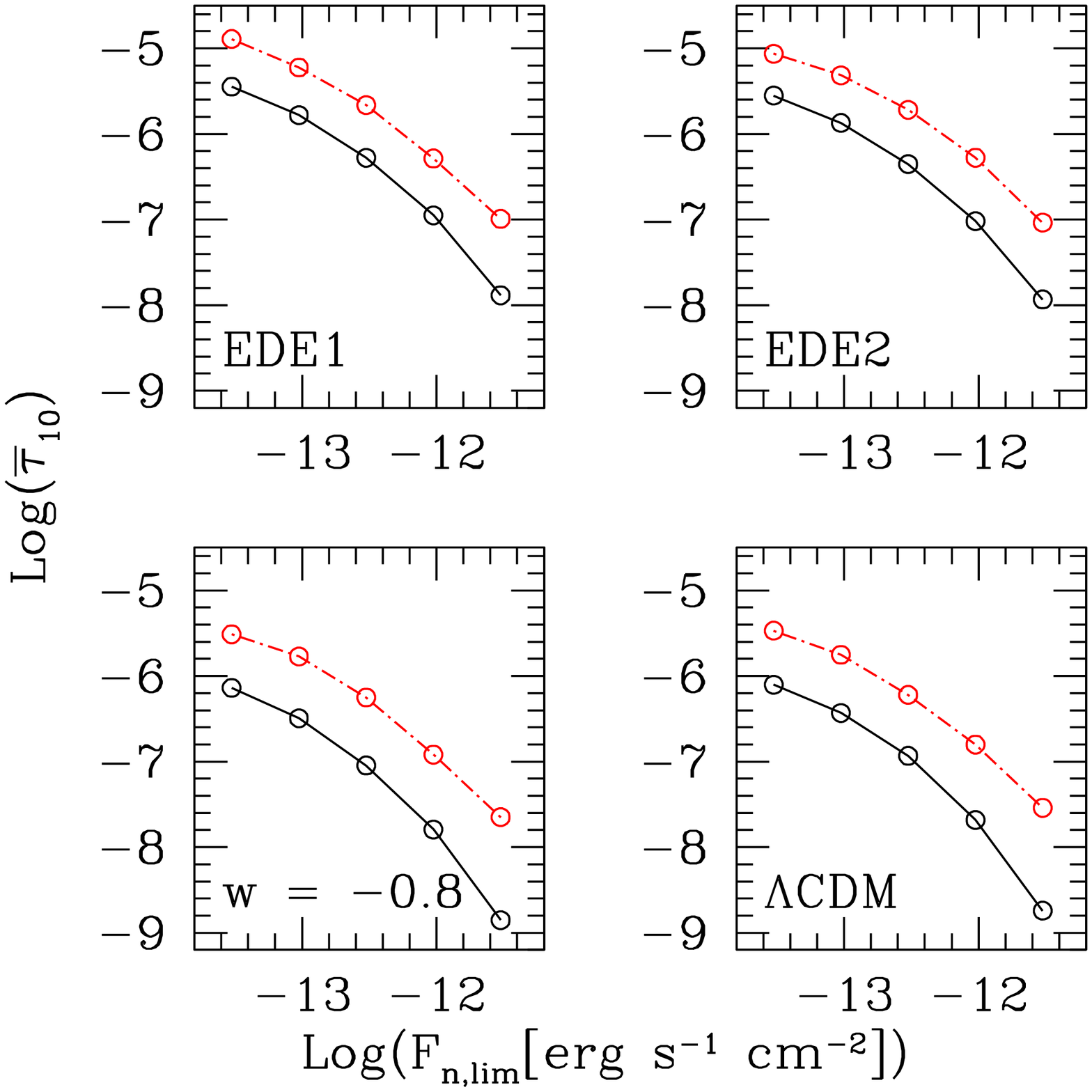}
\caption{The observed average optical depth for arcs with length to
width ratio larger than $d = 7.5$ (left panel) and $d = 10$ (right
panel) as a function of the nominal X-ray flux limit. Results are shown
for the four different cosmologies considered here. Black solid lines
represents results obtained by ignoring mergers processes, while broken
red lines are obtained taking them into account.}
\label{fig:tCounts}
\end{figure*}

Figure \ref{fig:tCounts} shows the average optical depth for
gravitational arcs with length to width ratio $d \ge 7.5$ and $d \ge 10$
predicted to be observed in a flux-limited X-ray cluster sample as a
function of the limiting flux. Results obtained both ignoring and taking
account of cluster mergers are shown. As noted in \cite{FE07.1} before,
mergers increase the optical depth by a factor between 2 and 3. The
present figure shows that this remains true for all flux limits
considered here even when X-ray selection effects are taken into
account. Moreover, we note that the slope of the $\bar{\tau} -
F_\mathrm{n,lim}$ relation tends to increase (decrease in absolute
value) towards low limiting fluxes. This is due to the fact that the
lensing efficiency drops above $z \approx 0.3$, and approaches zero
towards sufficiently high redshifts. Thus, if the flux limit is low
enough, the sample contains all the arcs that are produced and that
would be observed without selection effects.

\begin{figure*}[ht]
  \includegraphics[width=0.45\hsize]{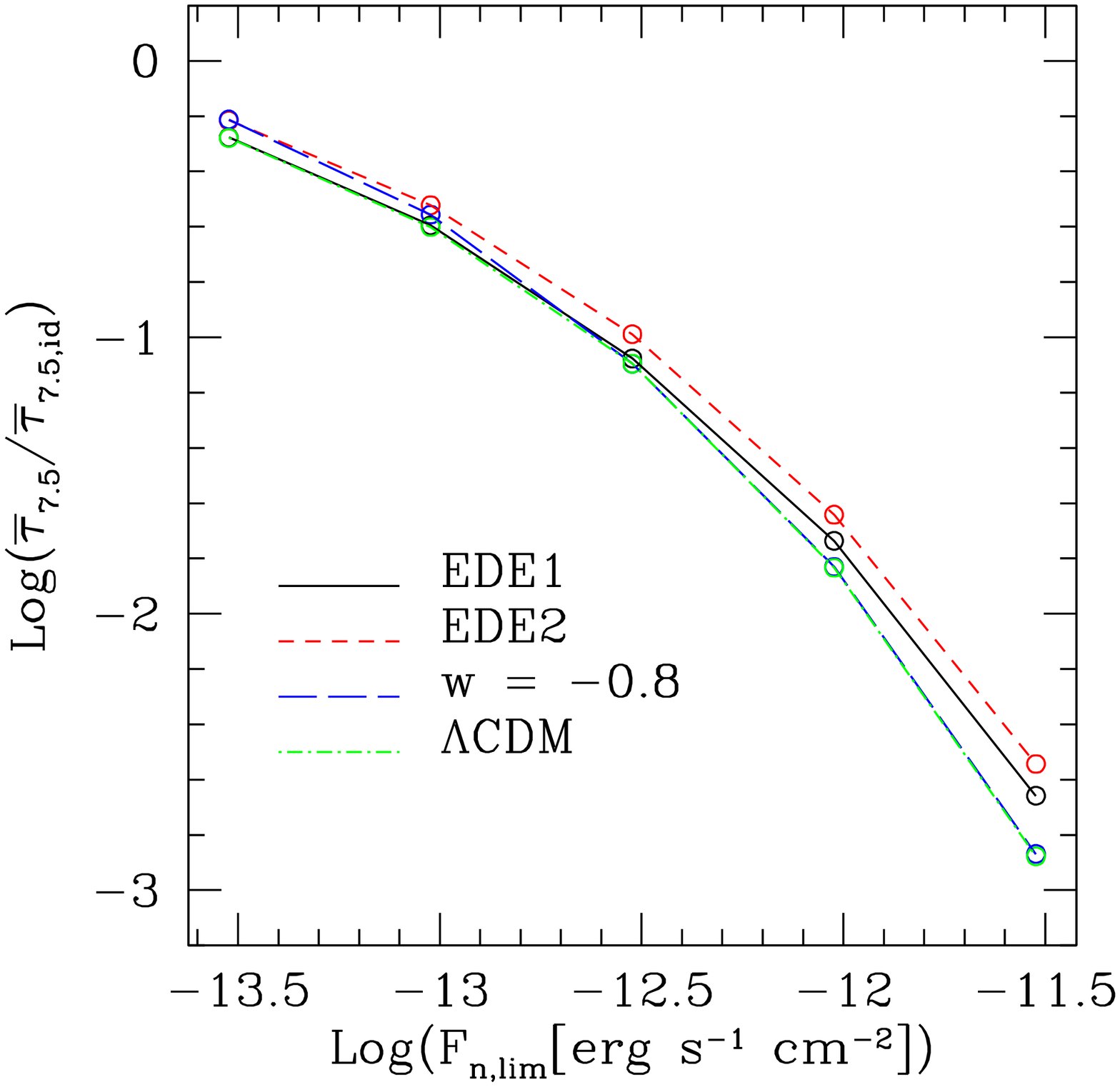}\hfill
  \includegraphics[width=0.45\hsize]{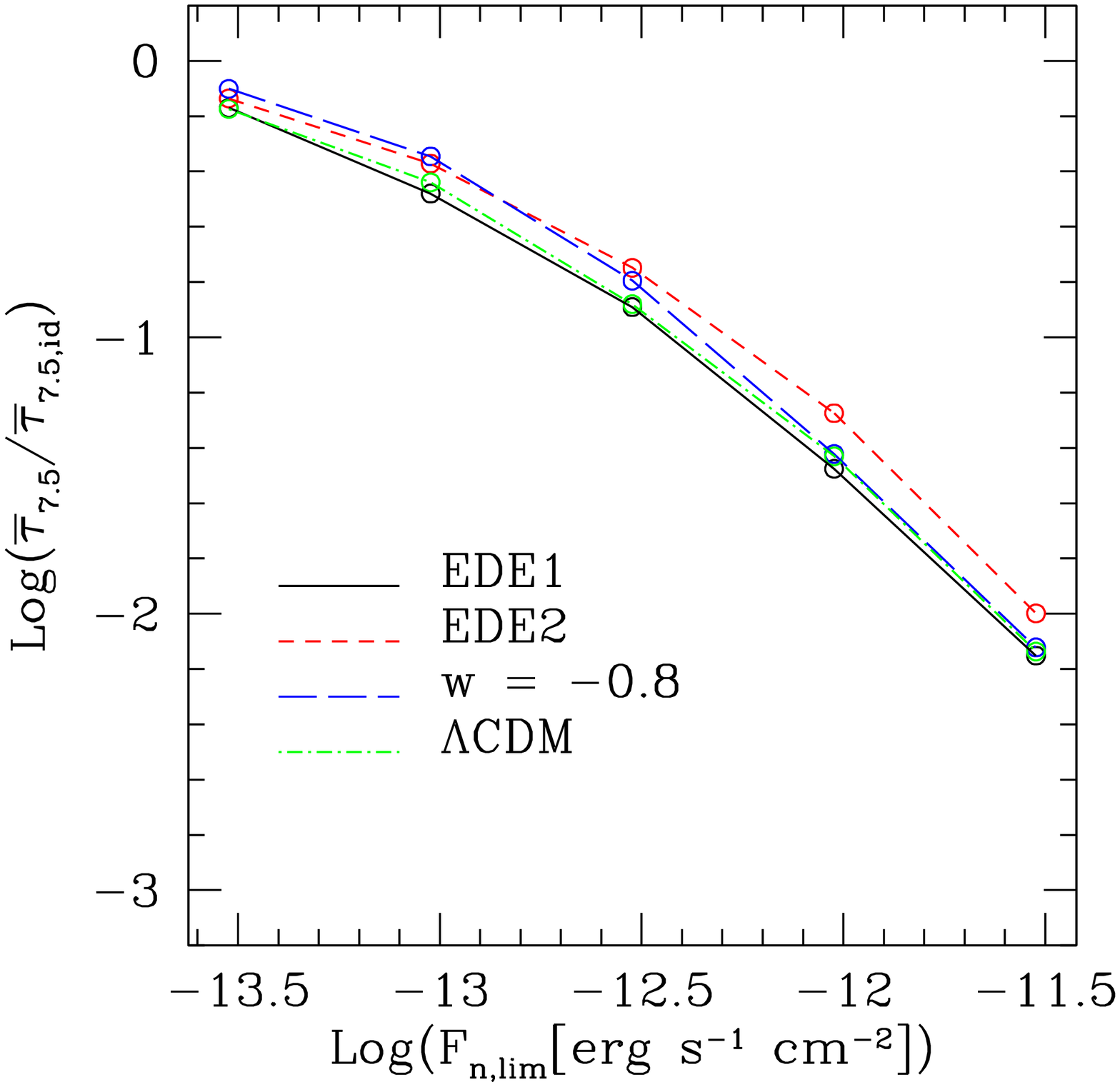}
  \includegraphics[width=0.45\hsize]{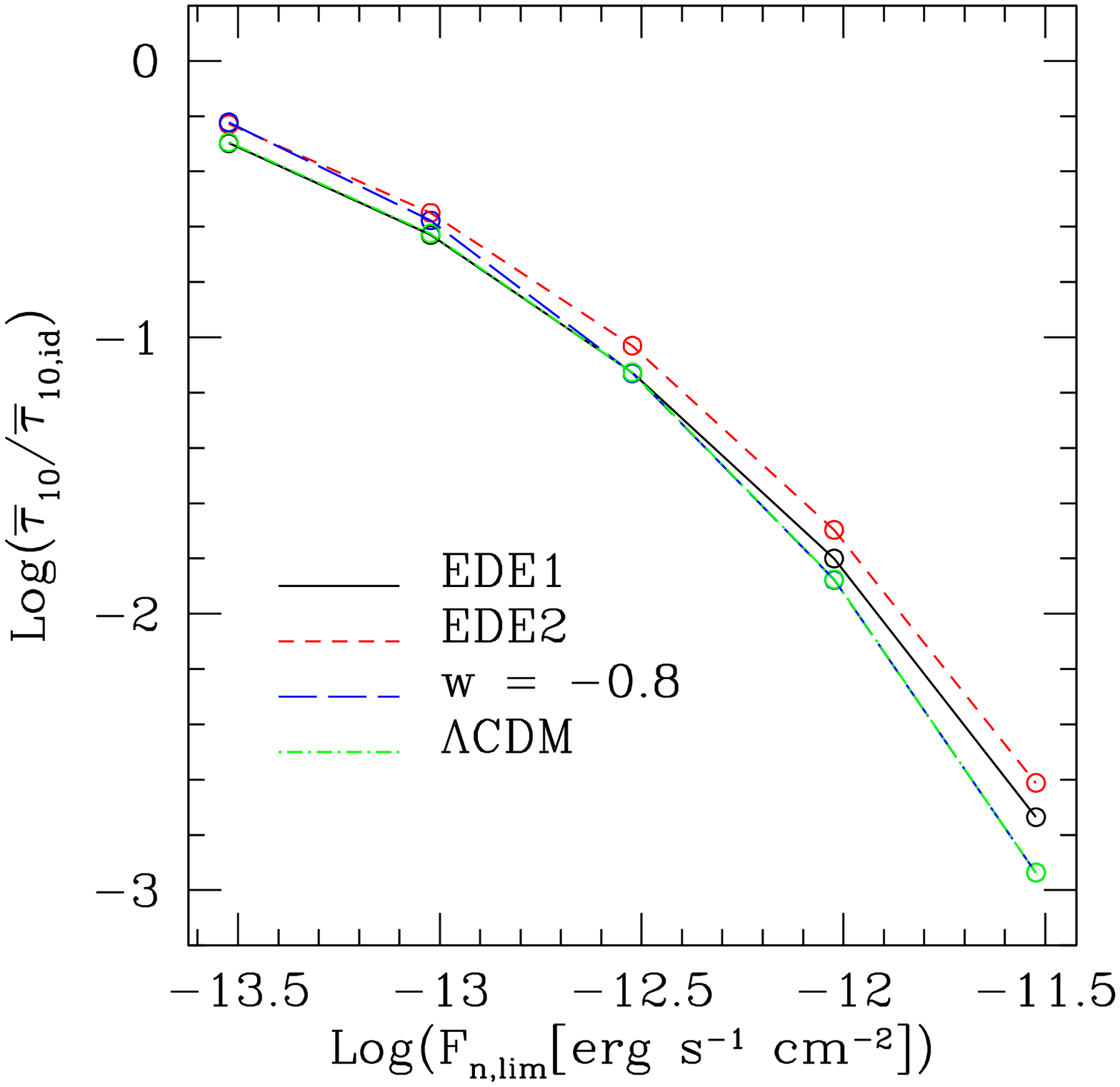}\hfill
  \includegraphics[width=0.45\hsize]{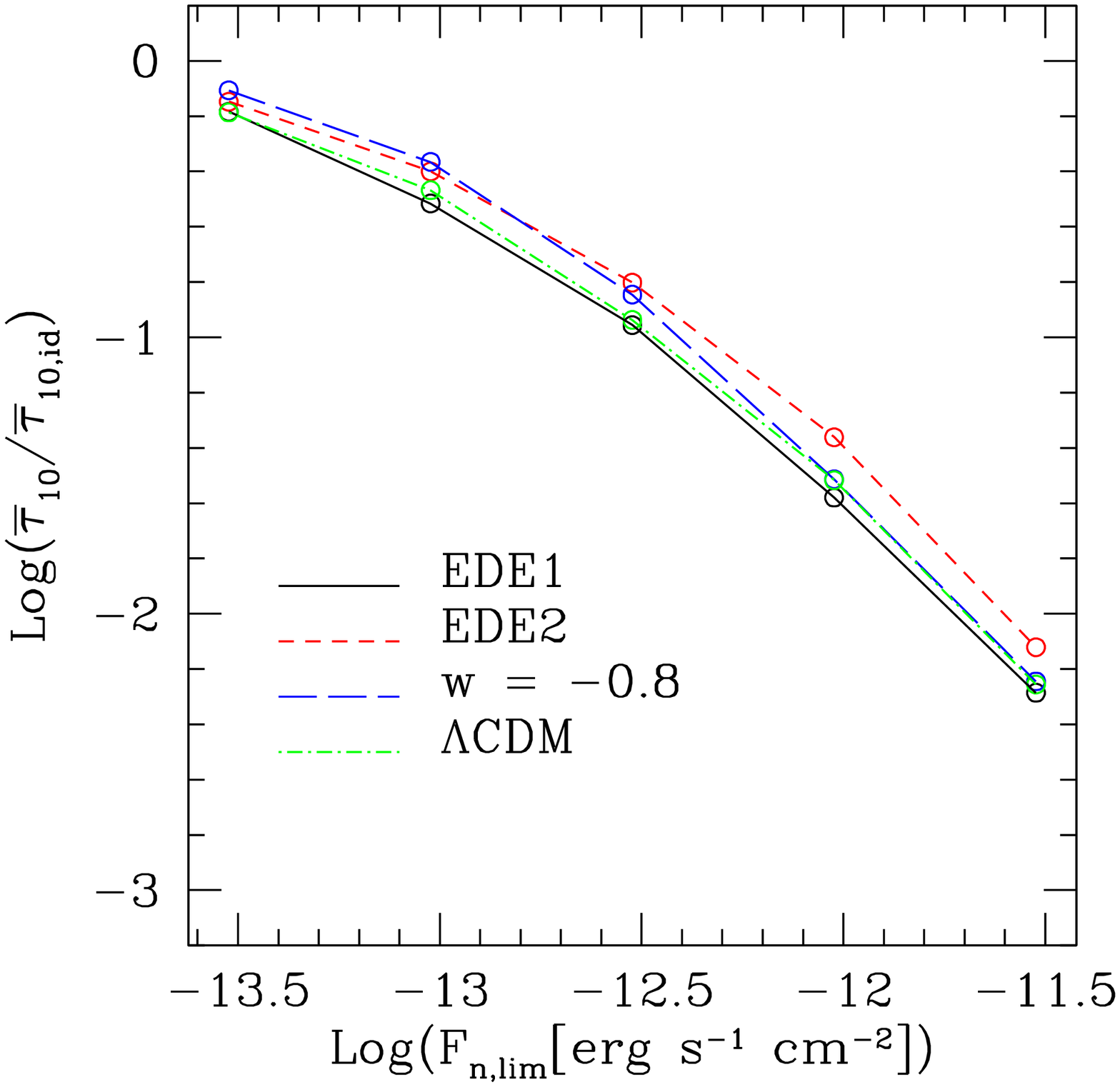}
\caption{The fraction of the optical depth for the formation of large
gravitational arcs, contributed by X-ray selected clusters in a
flux-limited sample, compared to the total optical depth. Different line
styles represent different cosmological models, as labelled in the plot.
The effects of cluster mergers are taken into account in the right panels
and ignored in the left panels. The upper and lower panels show results
for arcs with length-to-width ratios exceeding $d = 7.5$ and $d \ge 10$,
respectively.}
\label{fig:ratio}
\end{figure*}

Increasing the length-to-width threshold from $d = 7.5$ to $d = 10$
changes the absolute value of the average optical depth, but not its
qualitative behaviour as a function of the limiting flux.

A central result of our study is the ratio between the optical depth for
large gravitational arcs in an X-ray selected, flux-limited cluster
sample compared to the total, idealised optical depth.
Since the average optical
depth is related to the number of arcs by a constant factor (determined
by the total number density of background sources), this corresponds to
the ratio between the number of arcs expected to be seen in a
flux-limited sample of X-ray clusters, and the total number of arcs that
would be observable in absence of any X-ray selection effect. This ratio
is shown in the right and left panels of Fig.~\ref{fig:ratio},
accounting for and ignoring the transient merger boosts in temperature,
X-ray flux and lensing cross section, respectively.

We note that, when we include mergers, the fraction of the average
optical depth in an X-ray selected cluster sample increases with
increasing limiting flux. We attribute this to the facts that (i)
cluster mergers tend to enhance the lensing efficiency, which affects
both the X-ray selected and the total average optical depth, and (ii)
cluster mergers also enhance the clusters' temperature and the X-ray
flux, which only affect the X-ray selected optical depth. We also note
that there is a slight tendency for models with lower power-spectrum
normalisation $\sigma_8$ to have a larger fractional importance for the
observed average optical depth. For instance, the ratio between the
optical depth of X-ray selected clusters to the total optical depth is
systematically slightly larger for the model EDE2 ($\sigma_8 = 0.78$)
than for the model EDE1 ($\sigma_8 = 0.82$). However, this effect is
very small.

\section{Summary and conclusions}

We have studied here the influence of selection effects on the total
observed number of gravitational arcs in X-ray selected galaxy-cluster
samples, taking cluster mergers into account. 

To perform our study, we considered
the assembly history of a synthetic sample of galaxy clusters. We linked
the virial mass of the dark-matter cluster halo to the temperature of
the ICM using the virial relation (\ref{eqn:virial}). We then used the
analytic fitting formulae provided by \cite{RA02.1} for the boost in
temperature caused by merger processes between clusters
and substructures during the formation. Afterwards,
we used the publicly available software package \texttt{xspec} \citep{AR96.1}
to convert the (boosted and unboosted) temperature of the ICM into the
ideal flux (in front of the instrument) produced by each
individual object accounting for redshift, metal emission lines and
interstellar absorption. 
Then, using the response matrix of the PSPC detector
on-board the Rosat satellite, we transformed the ideal flux in a photon count
rate, also including observational effects such as realistic background
count rates and PSF convolution. We finally turned the count rates for each
object into a nominal flux, as defined in the construction of the
ROSAT-ESO Flux Limited X-ray cluster sample \textit{Reflex}.

We repeated this procedure to the four different cosmological models
previously used in \cite{FE07.1} for studying the lensing properties: A
$\Lambda$CDM model, a dark-energy universe with constant
equation-of-state parameter $w = -0.8$ for the dark-energy component,
and two early-dark energy models, whose relevance in terms of linear and
nonlinear structure formation has recently been pointed out.

As an intermediate, qualitative result, we analysed the total number of
clusters expected to be visible in X-ray selected cluster samples as a
function of the nominal flux limit. We obtain a significant difference
between cosmologies with early-dark energy and the model with constant
equation-of-state parameter, and also a significant difference due to
the introduction of the effect of cluster mergers. In particular, we
find that the qualitative redshift distribution of clusters observed in
the \textit{Reflex} sample can only be reproduced accounting for merger
boosts in temperature and luminosity. Moreover, early dark-energy models
seem to overpredict the total number of observed objects, although our
results are not precise at the low-mass end of the distribution, which
is irrelevant for strong lensing. Thus, our absolute numbers are likely
to be an overestimate. This may hint at a potentially very interesting
test for early-dark energy in particular, and for the dynamics of
quintessence models in general, and certainly warrants further
investigation in the future.

Finally, our main results are determined by the combination of the
nominal X-ray flux selection with the observed strong-lensing
statistics. We find that cluster mergers enhance the average observed
optical depth by factors between 2 and 3 for all limiting fluxes
considered here. We also confirm the result \cite{FE07.1} obtained
ignoring any X-ray selection effects, that the different
structure-formation history in early-dark energy models causes the
lensing efficiency to increase by a factor of $\approx 3$ compared to
models with constant equation-of-state parameter for the dark energy.
This remains true for all limiting fluxes, and we see that, for
instance, the same lensing efficiency reached in a $\Lambda$CDM model
with the help of cluster mergers is reached ignoring mergers in models
with early-dark energy, because of the higher cluster density at
moderate and high redshifts. Moreover, the slope of the
$\bar{\tau}_\mathrm{d} - F_\mathrm{n,lim}$ relation tends to flatten
towards lower limiting fluxes, indicating that we are approaching the
total average optical depth.

In an earlier study, \cite{ME05.1} addressed the difference between the
total strong-lensing optical depths of cluster populations in
cosmological models with different types of dark energy, finding that
dynamical dark energy tends to increase the optical depth for giant arcs
compared to the standard $\Lambda$CDM. While this seems to be a generic
prediction for dark-energy models, they used numerical simulations
ignoring X-ray selection effects, while we semi-analytically model the
cluster population, its lensing properties and selection issues. We
extend our method to models with an early dark-energy contribution.

We also find that the ratio of the flux-limited to the total average
optical depth is larger when we consider the effect of cluster mergers
than when we ignore it. This is due to the fact that the increment in
the lensing efficiency affects both the flux limited and the ideal
optical depth, while the boost in temperature and luminosity affects
only the former.

We carried out these calculations for gravitational arcs with
length-to-width ratios $d \ge 7.5$ and $d \ge 10$. We find an (expected)
difference in the absolute value of the average optical depth, while the
trend with the limiting X-ray flux is qualitatively unchanged.

Predicting the number of long and thin gravitational arcs to be observed
in X-ray selected cluster samples and in different cosmological models
will be very useful in the near future. Forthcoming strong lensing
surveys \citep{CA07.1} and the development of automatic detection
algorithms for strong lensing features \citep{LE04.1,HO05.3,SE06.1} will
allow to place further constraints on the dynamics of structure
formation in a universe dominated by dark energy.

\section*{Acknowledgements}

We are indebted to E. Puchwein, E. Rasia and A. Morandi for clarifying
many aspects of the \texttt{xspec} software package. Part of the work
has been performed under the Project HPC-EUROPA (RII3-CT-2003-506079),
with the support of the European Community - Research Infrastructure
Action under the FP6 ``Structuring the European Research Area''
Programme. This work was supported by the Collaborative Research Centre
SFB 439 of the \emph{Deutsche Forschungsgemeinschaft} and by the German
Academic Exchange Service (DAAD) under the \emph{Vigoni} programme. We
are grateful to M. Meneghetti and L. Moscardini for their support and
collaboration as well as for interesting and fruitful discussions. We also
thank the anonymous referee for useful remarks that helped
improving the presentation of this work.

\bibliographystyle{aa}
\bibliography{./master}

\end{document}